\documentclass[prb,twocolumn,showpacs,showkeys,amssymb]{revtex4}

\usepackage{graphicx}% Include figure files
\usepackage{dcolumn}% Align table columns on decimal point
\usepackage{bm}% bold math

\begin{document}

%\tightenlines

\title{Electronic Raman scattering in quantum dots revisited}
\author{Alain Delgado$^a$, Augusto Gonzalez$^b$, and D.J. Lockwood$^c$
\footnote{Corresponding author. email: David.Lockwood@nrc-cnrc.gc.ca}}
\affiliation{$^a$ Centro de Aplicaciones Tecnologicas y Desarrollo Nuclear,
 Calle 30 No 502, Miramar,Ciudad Habana, C.P. 11300, Cuba\\
 $^b$ Instituto de Cibernetica, Matematica
 y Fisica, Calle E 309, Vedado, Ciudad Habana, Cuba\\
 $^c$ Institute for Microstructural Sciences, National Research Council, 
 Ottawa, Canada K1A 0R6}

\begin{abstract}
We present theoretical results concerning inelastic light (Raman) scattering from 
semiconductor quantum dots. The characteristics of each dot state (whether it is a
collective or single-particle excitation, its multipolarity, and its spin) are 
determined independently of the Raman spectrum, in such a way that common beliefs
used for level assignments in experimental spectra can be tested. We explore the
usefulness of below band gap excitation and an external magnetic field to identify
charge and spin excited states of a collective or single-particle nature.
\end{abstract}


\maketitle

\section{Introduction}

Raman scattering in semiconductor structures was devised more than twenty years ago 
by Burstein et. al. as a powerful tool for the identification of electronic excitations \cite{Burstein}. To the
best of our knowledge, experiments on Raman scattering in quantum dots were performed
mainly before 1998. \cite{Lockwood1,Heitmann1,Lockwood2} The lack of theoretical 
calculations for the relatively large dots used in the experiments (dozens of electrons
per dot) made the experimental results less conclusive.

A second handicap of the experiments reported in Refs.
[\onlinecite{Lockwood1,Heitmann1,Lockwood2}] is related to the fact that they explored
the conceptually difficult resonant regime, where the incident photon energy is in
resonance with an electronic state in the conduction band. In contrast, the existing
(qualitative) theory of Raman scattering is expected to be valid only well away from
resonance \cite{ILS1}. It is usually called the off resonance approximation (ORA), and
is well known for missing out the single-particle peaks in the Raman spectrum
\cite{Sarma}, which are particularly important in the resonant regime. In Ref.
[\onlinecite{Heitmann1}], laser excitation energies 40 meV above band gap were used
to identify collective states. The physics of Raman scattering under these conditions 
is expected to be still more complex because of the sudden increase of level widths 
with the opening up of the channel for spontaneous emission of longitudinal optical
(LO) phonons \cite{phonons}. The positions of collective peaks were computed by
means of the ORA\cite{Barranco}, which constitutes a nice example of how one can
get reasonable results with a theory that does not work in this regime.

In the present paper, we review a set of theoretical results on Raman scattering
in relatively large quantum dots \cite{nuestro1,nuestro2,nuestro3,nuestro4}, which
were motivated by the experiments described in Refs.
[\onlinecite{Lockwood1,Heitmann1,Lockwood2}].

We explore the below band gap excitation regime to study how the ORA is reached. In
addition, it is shown that this regime is ideal for the identification of the collective
peaks and certain single-particle excitations (SPEs). Polarization rules for collective
states, following from the ORA, are tested. They are unexpectedly shown to work also
for the SPEs at zero magnetic field, and to break down in the presence of a field. 
Jump rules for the Raman peak intensities when the laser excitation energy approaches 
the band gap are also demonstrated. 

Raman spectra computed with resonant excitation (but with the laser energy below the
threshold for the creation of LO phonons) show extreme sensitivity to the excitation
energy, and a reinforcement of single-particle peaks.

We hope that our detailed findings will motivate more experimental work on the Raman
spectroscopy of quantum dots, which is regaining interest as a tool for studying small
(with less than 7 electrons) self-assembled dots\cite{Heitmann2}.

\section{Generalities}
\label{sec2}

\subsection{The experimentalist point of view}
\label{sec2A}

\begin{figure*}[ht]
\begin{center}
\includegraphics[width=1.0\linewidth,angle=0]{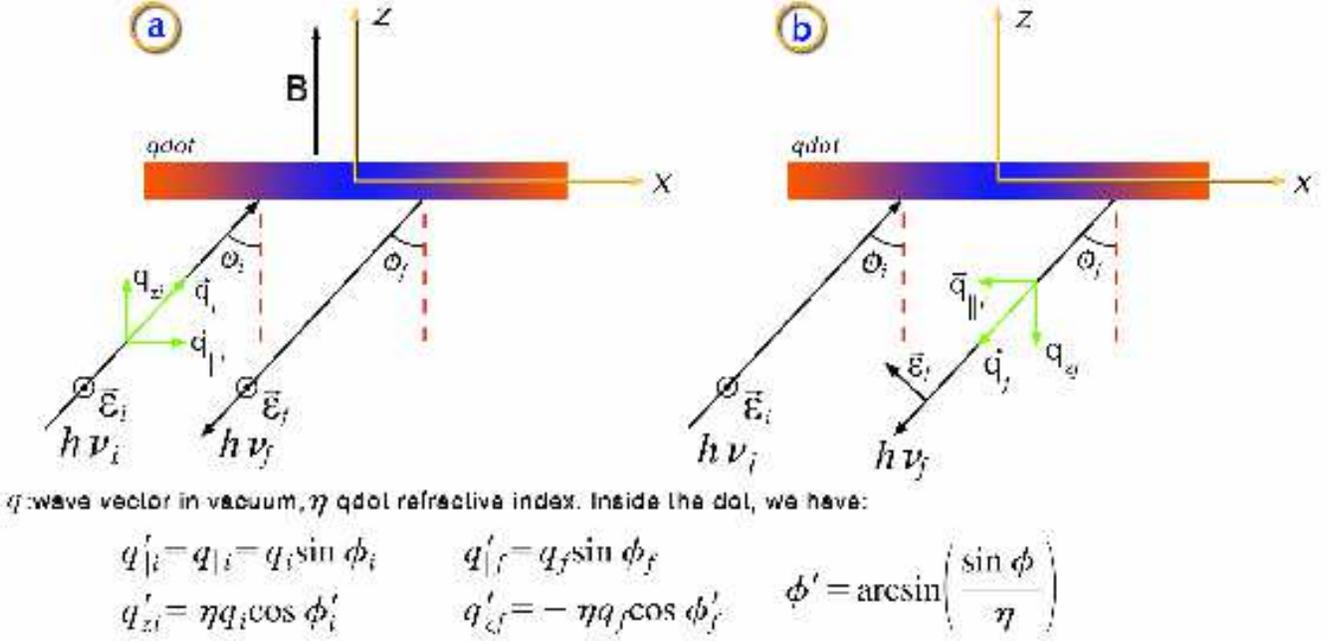}
\caption{\label{fig1} Geometry of the Raman experiment: (a) polarized, and
 (b) depolarized geometry.}
\end{center}
\end{figure*}

A typical experimental setup for Raman measurements is sketched in Fig. \ref{fig1}.
Usually, linear light polarization is used, although circular polarization would be
very convenient in certain situations. We take the incident electric field to be in the
plane of the dot, which is assumed quasi twodimensional.

The backscattering geometry, in which the incident and scattered light beams go through
the same optical fiber($\phi_i=\phi_f$), is also very common.

Two kinds of measurements are usually performed in order to separate charge and spin
excitations. In the so called polarized spectrum, (a), the final and incident electric
fields are parallel, $\vec E_i\parallel\vec E_f$. Whereas the depolarized spectrum is taken with $\vec E_i\perp\vec E_f$, (b). Notice that oblique incidence ($\phi_i,~
\phi_f\ne 0$) is used to excite multipole final states.

\subsection{The theorist point of view}
\label{sec2B}

The theoretical representation of a Raman process emerges from the second-order
perturbative result for the transition amplitude\cite{Loudon}, $A_{fi}$:

\begin{equation}
A_{fi}\sim \sum_{int} \frac{\langle f|H^+_{e-r}|int\rangle
 \langle int|H^-_{e-r}|i\rangle}{h\nu_i-(E_{int}-E_i)+i\Gamma_{int}}.
\label{eq1}
\end{equation}

\begin{figure}[ht]
\begin{center}
\includegraphics[width=1.0\linewidth,angle=0]{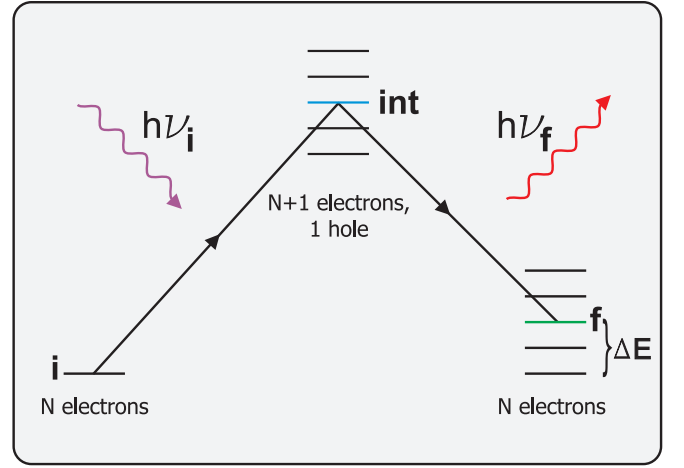}
\caption{\label{fig2} The quantum mechanical transition amplitude and its
 interpretation in terms of virtual transitions.}
\end{center}
\end{figure}

A schematic representation is given in Fig. \ref{fig2}. The Raman scattering process
can be viewed as the result of virtual transitions via all possible intermediate states.
In the experiments, the incident photon energy, $h\nu_i$, is close to the effective 
band gap, so that we can restrict the sum to intermediate states that contain an 
additional electron-hole pair.

In the experiments, the temperature, $T$, is usually below 2 K. We present calculations
at exactly $T=0$ K, in which only the ground state of the $N$-electron quantum dot is
initially populated. This will be our initial state, $|i\rangle$.

The intermediate electronic states, $|int\rangle$, as mentioned above, are states with
an additional electron-hole pair. This pair is (virtually) recombined leading to a 
photon with energy $h\nu_f$. The electronic system ends up in a final state, 
$|f\rangle$. The energy difference, $\Delta E_f=E_f-E_i$, is the Raman shift. Notice the width, $\Gamma_{int}$, entering the expression for the transition amplitude in Eq.
(\ref{eq1}). We will consider intermediate states in an energy window of 30 meV
above the band gap in order to neglect abrupt variations of $\Gamma_{int}$ due to
LO phonons. $\Gamma_{int}$ will be fixed phenomenologically to a constant value,
equal to 0.5 meV, which is due mainly to pair recombination.

\begin{figure*}[ht]
\begin{center}
\includegraphics[width=1.0\linewidth,angle=0]{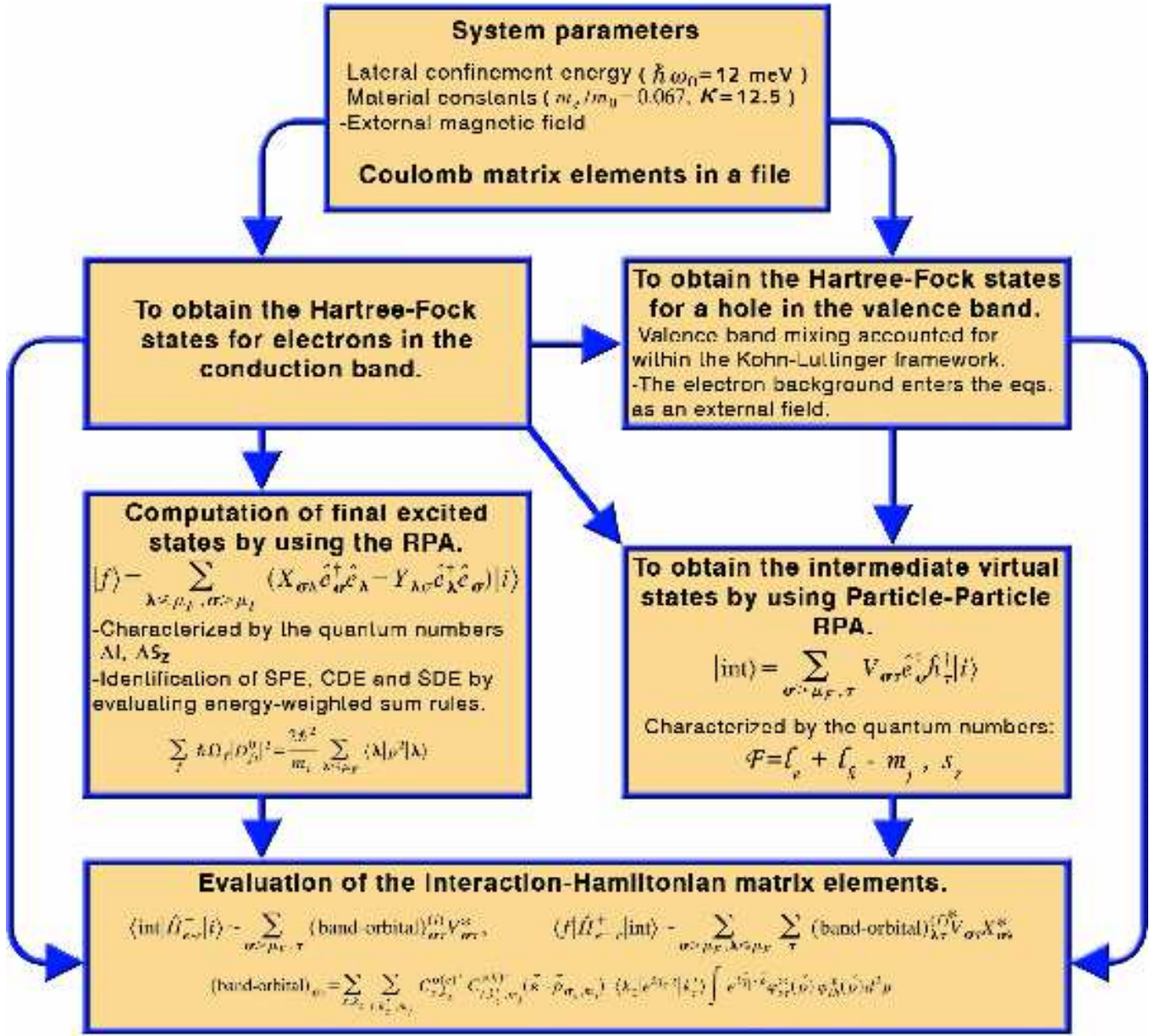}
\caption{\label{fig3} The calculation scheme used in the paper.}
\end{center}
\end{figure*}

From the transition amplitude one computes the cross section:

\begin{equation}
d\sigma\sim\sum_f |A_{fi}|^2 \delta(E_i+h\nu_i-E_f-h\nu_f).
\end{equation}

\noindent
Notice the energy conservation forced by the delta function, which implies that the 
Raman shift is equal to the photon energy loss. In our calculations, we replace the
delta function by a Lorentzian:

\begin{equation}
\delta(x)\approx \frac{\Gamma_f/\pi}{x^2+\Gamma_f^2},
\end{equation}

\noindent
where the width of final states will also be assumed constant, $\Gamma_f=0.1$ meV.
A 30 meV excitation window for final states will be used in order to avoid considering
phonon processes.

\subsection{Calculation machinery}
\label{sec2c}

The calculation scheme, explicit expressions, and explanations can be found by the
interested reader in papers [\onlinecite{nuestro2,nuestro5}]. We shall give in
this section an overview of the conceptual and numerical procedure with the aid of
Fig. \ref{fig3}.

The dot is modelled by a quantum well with hard walls in the $z$ direction (the 
growth direction of the heterostructure), and a soft harmonic confinement 
($\hbar\omega_0=12$ meV) in the $x,y$ plane. The basis functions used to describe
one-particle states in the dot are constructed as products of harmonic oscillator,
infinite well, and spin functions.

To save time in the calculation of many-electron wavefunctions, we computed the matrix elements of Coulomb interactions, $\langle\alpha,\beta|1/r|\gamma\delta\rangle$, where
$\alpha,\beta,\gamma,\delta$ are arbitrary one-particle states, and stored them in a computer file. This calculation takes around 7 days in a personal computer. The matrix
elements are loaded into the computer at the beginning of a calculation, allowing us to
solve the nonlinear integro-differential Hartree-Fock (HF) equations for 42 electrons
in a few minutes, or to compute all of the intermediate states entering a Raman process
(around 10 000) in a few days. The HF equations for holes include the electrostatic
field created by the background electrons in the dot, and take account of valence band 
mixing effects, as described by the Kohn-Luttinger Hamiltonian \cite{Bastard}. Up to
6 quantum-well sub-bands are considered in the HF equations for holes.

The final states of the Raman process are excited states of the $N$-electron system.
The intermediate states, on the other hand, are states with $N+1$ electrons and one
hole. Both kinds of states are computed by means of random-phase-approximation (RPA)
like ansatzs for the wavefunctions \cite{Ring}, as illustrated in Fig. \ref{fig3}.
It was already mentioned elsewhere \cite{nuestro2} that, in our opinion, the main
limitation of using these functions in the present context is not related to the
well-known lack of correlation effects in the RPA, but to the incomplete description of the
density of energy levels in intermediate and final states.

Final states are labelled by the quantum numbers $\Delta L$ and $\Delta S_z$, which
refer to changes (with respect to the ground state) in the total angular momentum
and total spin projections along the $z$ axis. Borrowing a terminology from Nuclear
Physics, we speak about monopole excitations when $\Delta L=0$, dipole excitations
when $\Delta L=\pm 1$, and quadrupole excitations when $\Delta L=\pm 2$. States are 
further classified by the degree of collectivity, computed with the help of 
energy-weighted sum rules \cite{Ring}. For charge monopole excitations, for example,
we have:

\begin{equation}
\sum_f \Delta E_f |D^0_{fi}|^2=\frac{2\hbar^2}{m_e}\sum_{\lambda\le\mu_F}
 \langle\lambda|\rho^2|\lambda\rangle.
\label{eq4}
\end{equation}

\noindent
The left hand side of this equation contains only many-particle magnitudes, i.e., 
the excitation energies and the matrix elements of the monopole operator between the
initial and final states, $D^0_{fi}$ (its explicit definition can be found in Ref.
\onlinecite{nuestro5}). The right hand side, however, can be evaluated in terms
of the occupied HF orbitals, $\lambda$. The magnitude $\rho$ is the cylindrical
coordinate in the plane. We will conventionally say that the final state $f$ is
a collective state if $\Delta E_f |D^0_{fi}|^2$ is greater than 5\% of the right
hand side of Eq. (\ref{eq4}). The same statement applies for other multipole 
excitations.

Concerning the spin quantum numbers, the formalism leads to ladders of final states
characterized by $\Delta S_z$. In order to discriminate between different total spin states within a single ladder, we compare the energy differences of states in
adjacent ladders with the corresponding Zeeman splitting. We talk about charge
excitations above the ground state when the total spin does not vary, $\Delta S=0$,
and spin excitations when $\Delta S=\pm 1$. The distinction between collective
and single-particle states is done on the basis of energy-weighted sum rules, which
can be written for any multipolarity and $\Delta S$. Collective excited states with 
$\Delta S=0$ are called charge density excitations (CDEs), whilst collective states
with $\Delta S=\pm 1$ are called spin density excitations (SDEs). The single-particle
excitations are further distinguished by their charge or spin nature. Thus, we will
refer to them as SPE(C) or SPE(S).

The intermediate states, on the other hand, are characterized by the spin $S_z$ of
the added electron, and the total pair angular momentum, ${\cal F}=l_e+l_h-m_j$.

Once the intermediate and final states of a Raman process are computed, the matrix
elements of the electron-radiation interaction Hamiltonian are evaluated in terms
of the coefficients entering the ansatz for the many-particle wavefunctions and the
coefficients of the HF expansion. A complete Raman calculation, which requires the
sum over all the intermediate and final states (previously stored into the 
computer), takes around one day in a personal computer.

\subsection{The ORA}
\label{sec2D}

The ORA is a limit in which the expression for the amplitude of a Raman process is
simplified. In this limit, only the initial and final state wavefunctions enter the
expression for $A_{fi}$. The sum over intermediate states disappears from it. This
means that, within the ORA, one can not explain phenomena such as intermediate
state resonances in Raman processes.

The explicit derivation of the ORA for Raman scattering in quantum dots can be found
in Ref. [\onlinecite{nuestro3}]. The assumptions for its derivation are basically two:
(i) The laser excitation energy is far away from any intermediate state energy, in
such a way that we can neglect the dependence of the denominator of Eq. (\ref{eq1}) on
$E_{int}$, and (ii) There is an energy window in the intermediate states (from 
$E_{gap}$ to approximately $E_{gap}+40$ meV) where the variations of $\Gamma_{int}$
can be neglected and the completeness relation is practically fulfilled:
$\sum_{int}' |int\rangle\langle int|\approx 1$. Under these assumptions, one can
write:

\begin{equation}
A_{fi}^{ORA}\sim\langle f|H^+_{e-r}H^-_{e-r}|i\rangle.
\end{equation}

\begin{figure*}[t]
\begin{center}
\includegraphics[width=.9\linewidth,angle=-90]{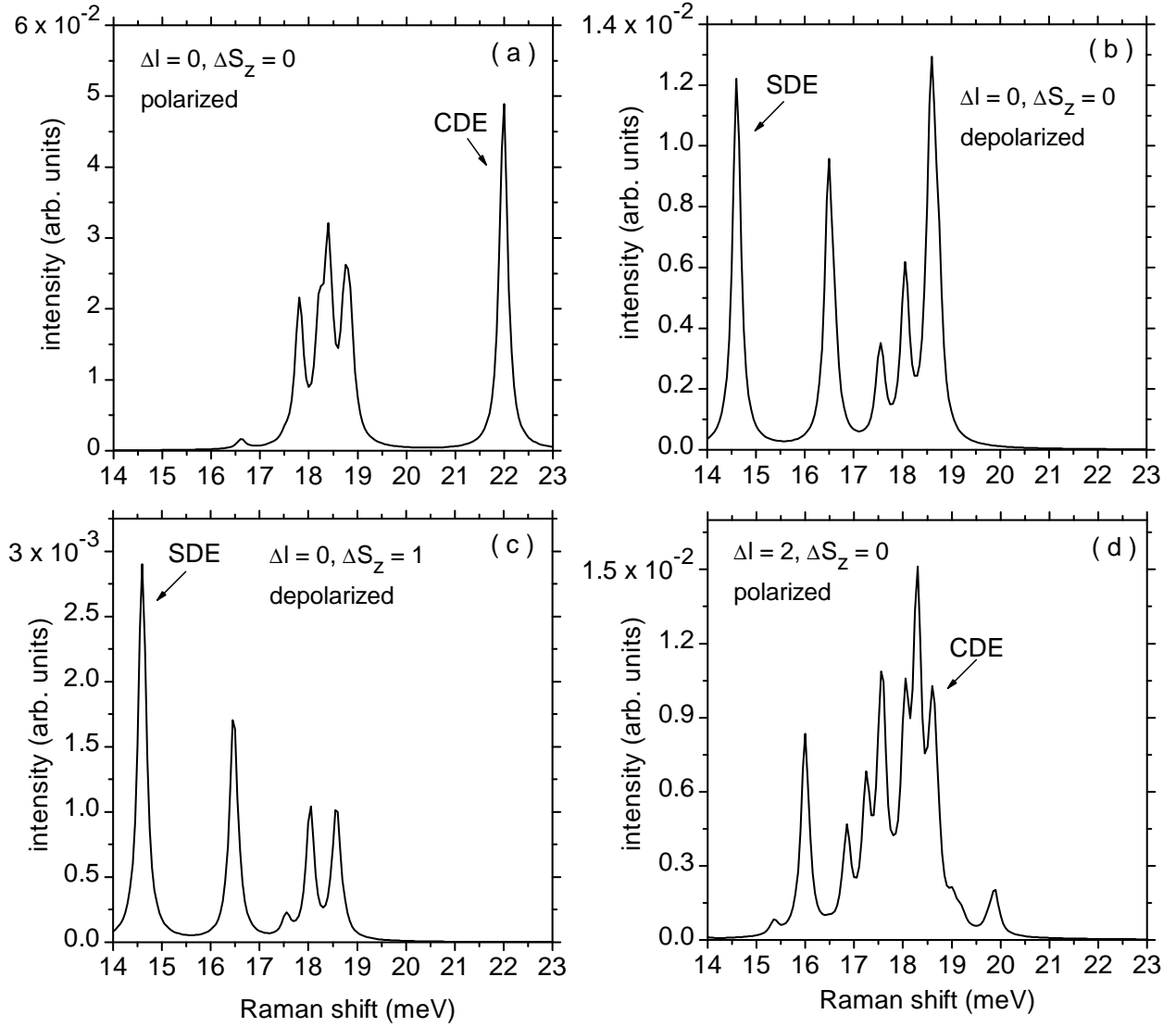}
\caption{\label{fig4} Calculated Raman spectra in different channels. The incident
 laser energy is $h\nu_i=E_{gap}-5$ meV.}
\end{center}
\end{figure*}

\noindent
Using the explicit expression for $H_{e-r}$, one arrives at the ORA formula:

\begin{widetext}
\begin{eqnarray}
A_{fi}^{ORA}&\sim&\sum_{\alpha,\alpha'}
 \langle\alpha| e^{i(\vec q_i-\vec q_f)\cdot \vec r}|\alpha'\rangle
 \left\{\frac{2}{3} (\vec \varepsilon_i\cdot\vec\varepsilon_f) 
 \left\langle f\left|~
 e^\dagger_{\alpha\uparrow} e_{\alpha'\uparrow}+
 e^\dagger_{\alpha\downarrow} e_{\alpha'\downarrow} 
 \right.\right|i\right\rangle\nonumber\\
&+&\frac{i}{3} (\vec \varepsilon_i\times\vec\varepsilon_f)\cdot 
 \left.\left\langle f\left|~
 \hat z~(e^\dagger_{\alpha\uparrow} e_{\alpha'\uparrow}-
 e^\dagger_{\alpha\downarrow} e_{\alpha'\downarrow})+
 (\hat x+i\hat y)~  e^\dagger_{\alpha\uparrow} e_{\alpha'\downarrow}+
 (\hat x-i\hat y)~ e^\dagger_{\alpha\downarrow} e_{\alpha'\uparrow}
    \right|i\right\rangle\right\}.
\label{eq6}
\end{eqnarray}
\end{widetext}

A few important conclusions may be derived from Eq. (\ref{eq6}). First, we notice that
only collective final states have a nonvanishing amplitude in this approximation. The
SPEs play no role in the Raman process. Indeed, by expanding the exponential in
Eq. (\ref{eq6}), one  can obtain an alternative expression for $A_{fi}^{ORA}$ in
terms of multipole operators \cite{nuestro5}. Only final states having nonzero
matrix elements of multipole operators will contribute to $A_{fi}^{ORA}$.

A second important conclusion is related to the spin selection rule for Raman
scattering. Notice that, in Eq. (\ref{eq6}), multipole operators that do not alter
the spin quantum numbers of the initial state are multiplied by the factor
$\vec \varepsilon_i\cdot\vec\varepsilon_f$. This means that peaks corresponding
to charge operators will appear in the polarized geometry. On the other hand, 
multipole operators that modify the spin are multiplied by the factor
$\vec \varepsilon_i\times\vec\varepsilon_f$ and, consequently, Raman peaks
corresponding to spin excitations are expected to be seen in the depolarized 
geometry.

\section{Results}
\label{sec3}

In the next two subsections, we present results for Raman intensities in quantum dots
for laser excitation energies below and above band gap. In the former case, no
experimental measurements have been performed so far. We expect weak Raman signals
in this regime, but there are also many advantages such as, for example, the absence of a luminescence background, a smooth dependence of peak intensities on the excitation
energy, etc \cite{nuestro2}. On the other hand, in the above band gap excitation regime, our
calculations are performed for an excitation window ranging from $E_{gap}$ to
$E_{gap}+30$ meV, where $E_{gap}$ is the effective quantum dot band gap. This is 
sometimes called the extreme resonance region.

\subsection{Raman spectra with below band gap excitation}
\label{sec3A}

The salient features of the Raman spectra in this excitation regime are summarized below.

\begin{figure}[ht]
\begin{center}
\includegraphics[width=.98\linewidth,angle=-90]{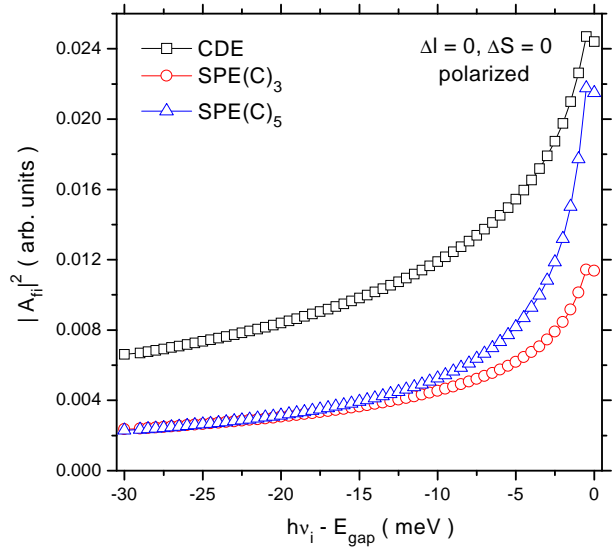}
\caption{\label{fig5} $|A_{fi}|^2$ for the monopolar CDE and two
 SPEs(C) with excitation energies 17.8 and 18.4 meV, respectively.}
\end{center}
\end{figure}

\subsubsection{Dominance of monopole peaks}
\label{sec3A1}

There are two reasons for the final state monopole excitations to be the dominant
peaks in the Raman spectrum. First, Raman scattering proceeds via the (virtual)
exchange of two photons. Selection rules dictate that the variation of the angular
momentum (with respect to the initial state) should be preferably $\Delta L=0$, 
$\pm 2$, etc. Second, the band-orbital factor in the matrix elements of $H_{e-r}$
(see Fig. \ref{fig3}) provides roughly a factor $(q r)^l$ every time a pair is created
or annihilated with a pair angular momentum $l$. As quantum dot dimensions are
typically $r\sim 100$ nm, \cite{Lockwood1,Heitmann1} and the laser wavelength is
$\lambda\sim 700$ nm, pairs with $l=0$ dominate the process.

On the other hand, final state spin excitations in which there is a spin flip with
respect to the initial state are depressed, as discussed elsewhere \cite{nuestro2}.
The Raman amplitude turns out to be proportional to the minority component of the
Kohn-Luttinger hole wavefunction. This means that monopole spin excitations in
which $\Delta S_z=0$ should be the dominant peaks in the depolarized geometry.

We show in Fig. \ref{fig4} some spectra in different angular momentum and spin
channels for the purpose of comparison. The laser energy is 5 meV below the band gap, and the incident (and backscattered) angle is equal to 20$^{\circ}$. Charge 
monopolar, charge quadrupolar, and spin monopolar peaks (with no spin flips)
exhibit comparable magnitudes. Dipolar (not shown) or spin-flip excitations lead
to Raman peak intensities one or two orders of magnitude lower than charge monopolar
peaks.

\begin{figure}[ht]
\begin{center}
\includegraphics[width=.98\linewidth,angle=0]{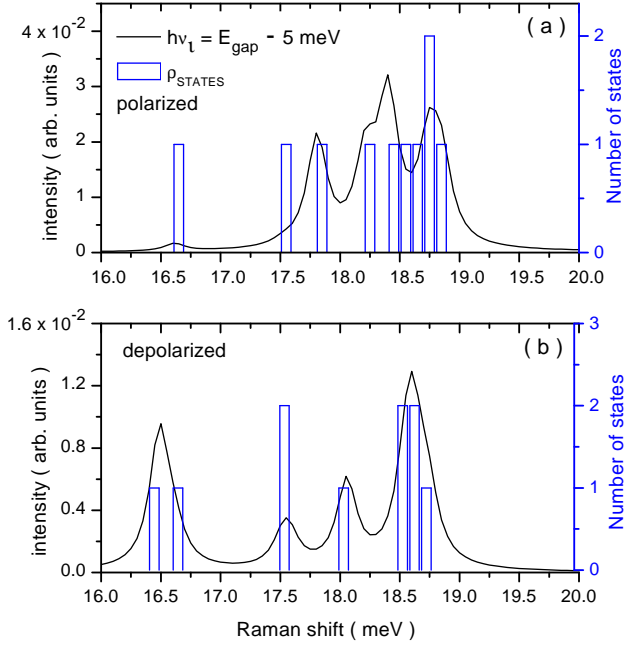}
\caption{\label{fig6} Polarized and depolarized monopolar Raman
 spectra and their comparison with the density of final state SPEs.}
\end{center}
\end{figure}

\subsubsection{Smooth dependence of peak intensities on the excitation energy}
\label{sec3A2}

The intensities of individual Raman peaks show a smooth dependence on $h\nu_i$
when the latter is below the band gap. This is simply understood from the 
Eq. (\ref{eq1}) for the transition amplitude. From the experimental point of view,
it is a nice feature. The identification of individual peaks could not be easier.

According to the ORA, as $h\nu_i$ moves away from the band gap, the collective states
should start dominating the Raman spectrum. In this way, we can identify the collective
and single-particle excitations (mainly the monopolar and quadrupolar ones) by 
varying $h\nu_i$. We show in Fig. \ref{fig5} the amplitude squared, $|A_{fi}|^2$, 
computed in the polarized geometry, for three charge monopolar final states. One of
them is the CDE, and the other two are SPEs with excitation energies 17.8 and 18.4
meV, respectively. The CDE already becomes the dominant peak when $h\nu_i$ is around 3
meV below the band gap.

\begin{figure}[ht]
\begin{center}
\includegraphics[width=1.\linewidth,angle=0]{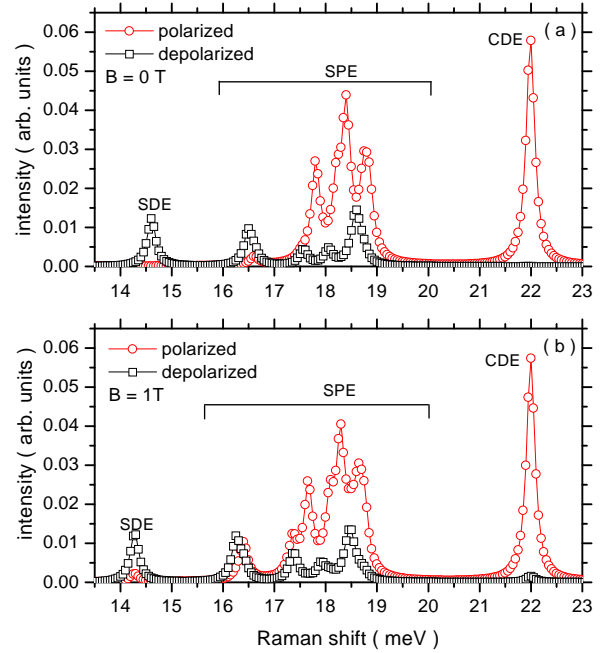}
\caption{\label{fig7} Polarized and depolarized monopolar Raman
 spectra at $B=0$ and 1 T.}
\end{center}
\end{figure}

\subsubsection{Correlation between the SPE Raman peaks and the density of final state
 energy levels}
\label{sec3A3}

It is natural to expect Raman peaks to be located there where there is an agglomeration
of final states \cite{Lockwood1}. In this paragraph, we take a step further and 
compare the polarized Raman spectrum with the density of final state SPEs(C), and the
depolarized spectrum with the density of SPEs(S). This is an attempt to test the spin
selection rules, derived from the ORA for the collective excitations, in the SPEs.

The comparison is given in Fig. \ref{fig6}, where we show results in the monopolar
channel. Unexpectedly, the correlation is high, particularly in the depolarized 
geometry. It is instructive to comment on a commonly held belief that the SPEs
appear in both polarized and depolarized spectra. Of course, it is true. But it
would be better to say that the SPEs(C) appear mainly in the polarized geometry, and 
the SPEs(S) mainly in the depolarized geometry. This statement is valid at zero
magnetic field.

\subsubsection{Breakdown of the polarization selection rules in a magnetic field}
\label{sec3A4}

In a magnetic field, the selection rules deduced from the ORA for the collective
states are no longer valid at excitation energies close to the band gap. The reason 
is the magnetic field dependence of energies and wavefunctions of intermediate states
entering the summation of Eq. (\ref{eq1}).
The situation is depicted in Fig. \ref{fig7}, where monopolar Raman spectra at $B=0$
and 1 T are drawn. The excitation energy is $h\nu_i=E_{gap}-2.5$ meV.

Let us define the polarization ratio of a single final state, $|f\rangle$, as the 
ratio of $|A_{fi}|^2$ in unfavorable and favorable geometries, i.e., 
$r=|A_{fi}({\rm unfavorable})|^2/|A_{fi}({\rm favorable})|^2$. By favorable we mean
the polarized geometry for a charge excitation, and the depolarized geometry for a spin 
excitation.

\begin{figure}[ht]
\begin{center}
\includegraphics[width=1.\linewidth,angle=0]{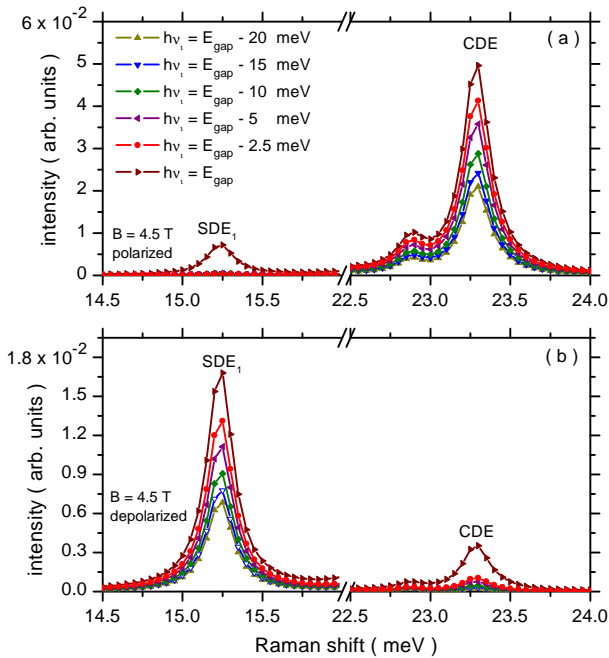}
\caption{\label{fig8} Polarized and depolarized monopolar Raman
 spectra at $B=4.5$ T as $h\nu_i$ approaches the band gap.}
\end{center}
\end{figure}

The polarization ratios for the collective states at $B=0$ are $3\times 10^{-4}$ for
the CDE, and $2\times 10^{-7}$ for the SDE. That is, an almost perfect fulfillment
of the ORA selection rules in spite of the fact that $h\nu_i$ is only 2.5 meV below band gap. At $B=1$ T, however, these numbers change to 0.03 and 0.17, respectively.

The fulfillment of the selection rules for the SPEs is not so evident in Fig. \ref{fig7},
because these states, with excitation energies between 16 and 19 meV, are not as 
easily distinguishable as the collective ones. From the data used to draw this figure,
we can compute $r$ for each SPE. The average polarization ratios at $B=0$ are 
$3\times 10^{-3}$ for the SPEs(C), and $2\times 10^{-4}$ for the SPEs(S). At $B=1$
T, these numbers become 0.26 and 1.1, respectively. That is, the intensities in
both geometries become comparable.

\subsubsection{Jump rule at the band gap}
\label{sec3A5}

Near the band gap, the intensities of Raman peaks follow an interesting behavior. Let 
us focus on the collective excitations and compare the intensities of the CDE and SDE 
peaks as a function of $h\nu_i$. The results for the monopolar spectra at $B=4.5$
T are shown in Fig. \ref{fig8}.

\begin{figure}[ht]
\begin{center}
\includegraphics[width=.95\linewidth,angle=-90]{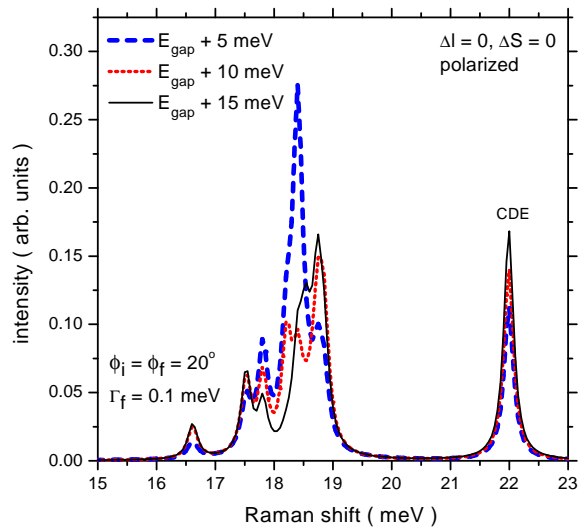}
\caption{\label{fig9} Monopolar Raman spectra above band gap in the
 polarized geometry.}
\end{center}
\end{figure}

We observe a smooth increase of peak intensity in the favorable geometry as $h\nu_i$
approaches the band gap. In the unfavorable geometry, however, the intensity remains
practically constant up to the moment when $h\nu_i$ reaches the band gap, where there
is a sudden variation. We call this phenomenon the ``jump rule''. SPEs also largely
follow this rule. 

\subsection{Raman spectra under resonant excitation}
\label{sec3B}

We return now to the $B=0$ case, and increase the laser energy to values above the
band gap. As mentioned above, we restrict $h\nu_i$ to the interval ($E_{gap},E_{gap}+
30$ meV) in order to avoid considering phonon effects.

Many of the properties discussed in the preceding section are still valid in the
present context. For example, the monopolar and quadrupolar peaks are the most
important peaks in the Raman spectrum, and the density of final state energy levels is 
correlated with the positions of the principal Raman lines.

In addition, there are certain new features, which characterize the resonant excitation
regime.

\begin{figure}[ht]
\begin{center}
\includegraphics[width=.95\linewidth,angle=0]{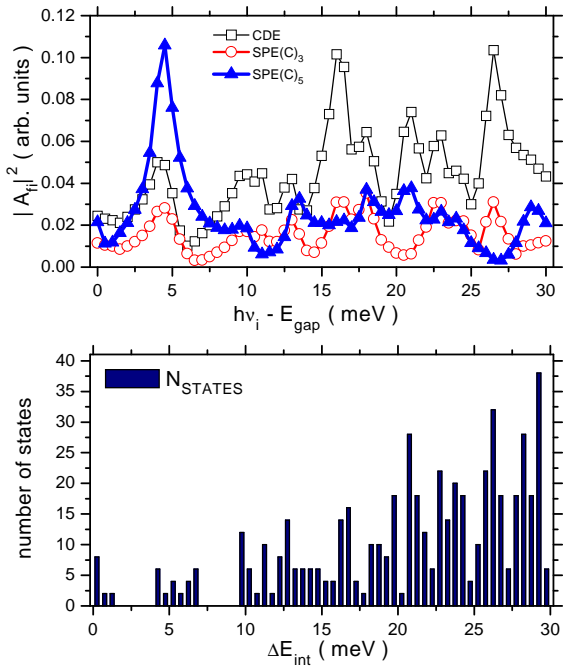}
\caption{\label{fig10} Upper panel: $|A_{fi}|^2$ for the same three charge
 monopolar states used in Fig. \ref{fig5} and $h\nu_i$ above band gap. Lower panel: Histogram
 of intermediate state energy levels.}
\end{center}
\end{figure}

\subsubsection{Dominance of SPEs}
\label{sec3B1}

Raman peaks related to SPEs experience a noticeable increase of intensity under 
resonant excitation. Qualitatively speaking, one can say that Raman scattering proceeds
through a single intermediate state (the one exactly at resonance), which (virtually)
decays indiscriminately to the collective final state or to the SPEs. As there is a large number of SPEs, and they are packed in groups, the intensities of the corresponding peaks can be high.

This statement is illustrated in Fig. \ref{fig9}, where three spectra corresponding
to $h\nu_i=E_{gap}+5$ meV, $E_{gap}+10$ meV, and $E_{gap}+15$ meV, respectively, are
shown. In addition, we observe a non-monotonous dependence of peak intensities on
$h\nu_i$. This is a consequence of the fact that, when $h\nu_i$ is varied, a different
intermediate state comes into resonance.

\subsubsection{Correlation between the Raman intensities of individual states
 and the density of intermediate state energy levels}
\label{sec3B2}

Here, we show how by monitoring the intensities of individual peaks as
a function of $h\nu_i$ one can obtain information about the density of energy levels
in intermediate states, at least at low excitation energies above the band gap.

In Fig. \ref{fig10}, we follow the same final states used in Fig. \ref{fig5}, and
compare the corresponding $|A_{fi}|^2$ for $h\nu_i$ in the interval ($E_{gap},E_{gap}+
30$ meV) with the density of energy levels. 

The first peak of $|A_{fi}|^2$  near 5 meV above the band gap signals the beginning
of a group of energy levels. A second structure is seen near 10 meV above the band
gap, where there is also a threshold. At higher excitation energies there is still some
correlation between peaks in $|A_{fi}|^2$ and peaks in the density of intermediate state energy levels. 

\begin{figure}[ht]
\begin{center}
\includegraphics[width=.9\linewidth,angle=-90]{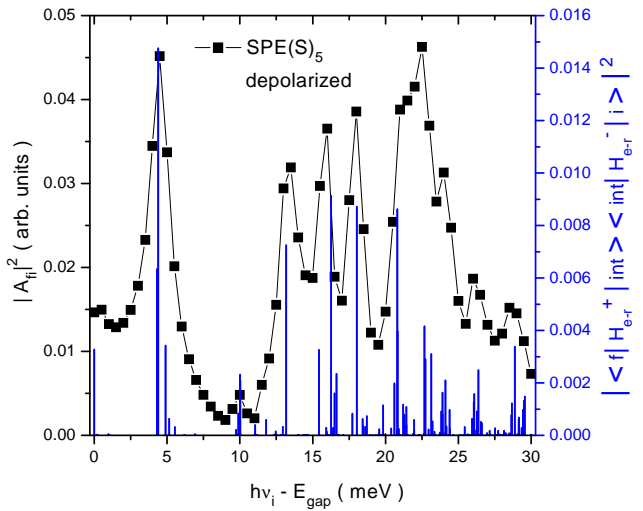}
\caption{\label{fig11} $|A_{fi}|^2$ for the spin monopolar state
 with excitation energy $\Delta E_f=18$ meV and the contribution of each
 individual intermediate state to the sum.}
\end{center}
\end{figure}

\subsubsection{Absence of interference effects}
\label{sec3B3}

We now evaluate the contribution of the different intermediate states entering the
summation of Eq. (\ref{eq1}) in resonant Raman scattering. The postulate is that the 
qualitative picture sketched in Sec. \ref{sec3B1} is correct: only those intermediate
states whose energies are very close to $h\nu_i$ contribute to the Raman amplitude.
To verify this, we performed different calculations of $A_{fi}$, restricting the sum 
over intermediate states to the window ($h\nu_i-\delta E$,$h\nu_i+\delta E$). There is
qualitatively no change in the spectrum when $\delta E$ is reduced from 10 down
to 2 meV.

We compare in Fig. \ref{fig11} the magnitude $|A_{fi}|^2$ for the spin monopolar peak
with $\Delta E_f=18$ meV with the individual contributions of each intermediate state
to the sum. No strong cancellation effects nor strong cooperation effects are
apparent. That is, interference effects in $A_{fi}$ are weak under resonant excitation.

\subsubsection{Selection rules in a magnetic field}
\label{sec3B4}

The polarization selection rules that follow from the ORA may be tested for excitation
energies above the band gap. Surprisingly, they shown to be very well
obeyed at zero magnetic field, and break down in a magnetic field. The polarization
ratio exhibits a strong dependence on $h\nu_i$ and on the applied field.  In quality 
of example, let us consider the monopole SDE. At zero field, $r$ is practically zero 
($10^{-7}$ when $h\nu_i=E_{gap}+5$ meV). But at $B=1$ T, $r$ varies from 1.1 at the
band gap to 0.1 for $h\nu_i=E_{gap}+2.5$ meV.

\section{Conclusions}
\label{sec4}

These theoretical calculations of the Raman spectrum of a many-electron quantum dot
in zero and non zero applied magnetic field have revealed intriguing new features
and a great sensitivity to the excitation energy. 

For excitation energies below the band gap, we found Raman spectra dominated by the
collective excitations already when $h\nu_i=E_{gap}-5$ meV. The peak intensities
depend smoothly on the excitation energy. In zero field, the simple-minded
polarization selection rules -- CDE polarized/SDE depolarized -- are very well
obeyed even by the SPEs. This means that one can obtain information about the
density of final state SPEs from the Raman spectra. In an applied magnetic field,
these basic selection rules break down, particularly when $h\nu_i$ is increased up 
to the band gap, where the Raman intensities of the collective excitations in their
``forbidden'' geometries suddenly increase to appreciable values. This we termed
the Raman intensity ``jump rule'', and may help to determine the charge or spin
character of a single excitation.

A quite different situation is encountered for excitation at energies above the
band gap. The Raman peak intensities fluctuate considerably as the excitation
energy sweeps through resonance with intermediate states. In this regime, the
SPEs dominate the Raman spectrum, the Raman intensities of individual peaks 
correlate well with the density of intermediate state energy levels, and there is
an absence of interference effects.

It would be interesting now to explore experimentally these new features of the
electronic Raman spectrum of many-electron quantum dots.

\begin{acknowledgments}
Part of this work was performed at the Institute for Microstructural Sciences (IMS),
National Research Council, Ottawa. A.D. acknowledges the hospitality and support of
IMS.
\end{acknowledgments}

\end{document}